\documentclass[conference]{IEEEtran}
\IEEEoverridecommandlockouts
\usepackage{cite}
\usepackage{amsmath,amssymb,amsfonts}
\usepackage{graphicx}
\usepackage{textcomp}
\usepackage{xcolor}
\usepackage{bm}
\usepackage{amsmath}
\usepackage{amssymb}
\usepackage{amsthm}
\usepackage{mathrsfs}
\usepackage{enumerate}
\usepackage{multirow}
\usepackage{color}
\usepackage{threeparttable}
\usepackage{subfigure}
\usepackage{caption}
\usepackage{float}
\usepackage{makecell}  
\usepackage[square, comma, sort&compress, numbers]{natbib}
\usepackage[pagebackref=false,breaklinks=true,letterpaper=true,colorlinks,bookmarks=false]{hyperref}
\usepackage{cleveref}

\usepackage{times}
\usepackage{breakurl}
\usepackage{array}
\usepackage{verbatim}
\usepackage{algorithm}
\usepackage{bbding}
\usepackage{algpseudocode}
\usepackage{lettrine}

\def\BibTeX{{\rm B\kern-.05em{\sc i\kern-.025em b}\kern-.08em
    T\kern-.1667em\lower.7ex\hbox{E}\kern-.125emX}}
\begin{document}
\title{S\textsuperscript{3}TU-Net: Structured Convolution and Superpixel Transformer for Lung Nodule Segmentation\\
\thanks{}
\author{
    \IEEEauthorblockN{Yuke Wu\textsuperscript{1}, Xiang Liu\textsuperscript{1*}, Yunyu Shi\textsuperscript{1}, Xinyi Chen\textsuperscript{1}, Zhenglei Wang\textsuperscript{2}, YuQing Xu\textsuperscript{3}, Shuo Hong Wang\textsuperscript{4}} 
    \IEEEauthorblockA{\textsuperscript{1}School of Electronic and Electrical Engineering, Shanghai University of Engineering Technology, Shanghai, China \\
    Email: M325122219@sues.edu.cn, xliu@sues.edu.cn, yunyushi@sues.edu.cn, c2257873708@163.com}
    \IEEEauthorblockA{\textsuperscript{2}Department of Medical Imaging, Shanghai Electric Power Hospital, Shanghai, China \\
    Email: hanqi\_willis@163.com}
    \IEEEauthorblockA{\textsuperscript{3}Department of Cloud Computing, Shanghai IDEAL INFORMATION Industry Co., LTD, Shanghai, China \\
    Email: xuyuqing.sh@chinatelecom.cn}
    \IEEEauthorblockA{\textsuperscript{4}Department of Molecular and Cellular Biology, Center for Brain Science, Harvard University, Cambridge, MA, USA \\
    Email: wangsh@fas.harvard.edu}
}
}
\maketitle

\begin{abstract}
The irregular and challenging characteristics of lung adenocarcinoma nodules in computed tomography (CT) images complicate staging diagnosis, making accurate segmentation critical for clinicians to extract detailed lesion information. In this study, we propose a segmentation model, S\textsuperscript{3}TU-Net, which integrates multi-dimensional spatial connectors and a superpixel-based visual transformer. S\textsuperscript{3}TU-Net is built on a multi-view CNN-Transformer hybrid architecture, incorporating superpixel algorithms, structured weighting, and spatial shifting techniques to achieve superior segmentation performance. The model leverages structured convolution blocks (DWF-Conv/D\textsuperscript{2}BR-Conv) to extract multi-scale local features while mitigating overfitting. 
To enhance multi-scale feature fusion, we introduce the S\textsuperscript{2}-MLP Link, integrating spatial shifting and attention mechanisms at the skip connections. Additionally, the residual-based superpixel visual transformer (RM-SViT) effectively merges global and local features by employing sparse correlation learning and multi-branch attention to capture long-range dependencies, with residual connections enhancing stability and computational efficiency. Experimental results on the LIDC-IDRI dataset demonstrate that S\textsuperscript{3}TU-Net achieves a DSC, precision, and IoU of 89.04\%, 90.73\%, and 90.70\%, respectively. Compared to recent methods, S\textsuperscript{3}TU-Net improves DSC by 4.52\% and sensitivity by 3.16\%, with other metrics showing an approximate 2\% increase. In addition to comparison and ablation studies, we validated the generalization ability of our model on the EPDB private dataset, achieving a DSC of 86.40\%.

\end{abstract}


\section{Introduction}
\begin{figure*}[t]  
    \centering
    \includegraphics[width=1.0\linewidth]{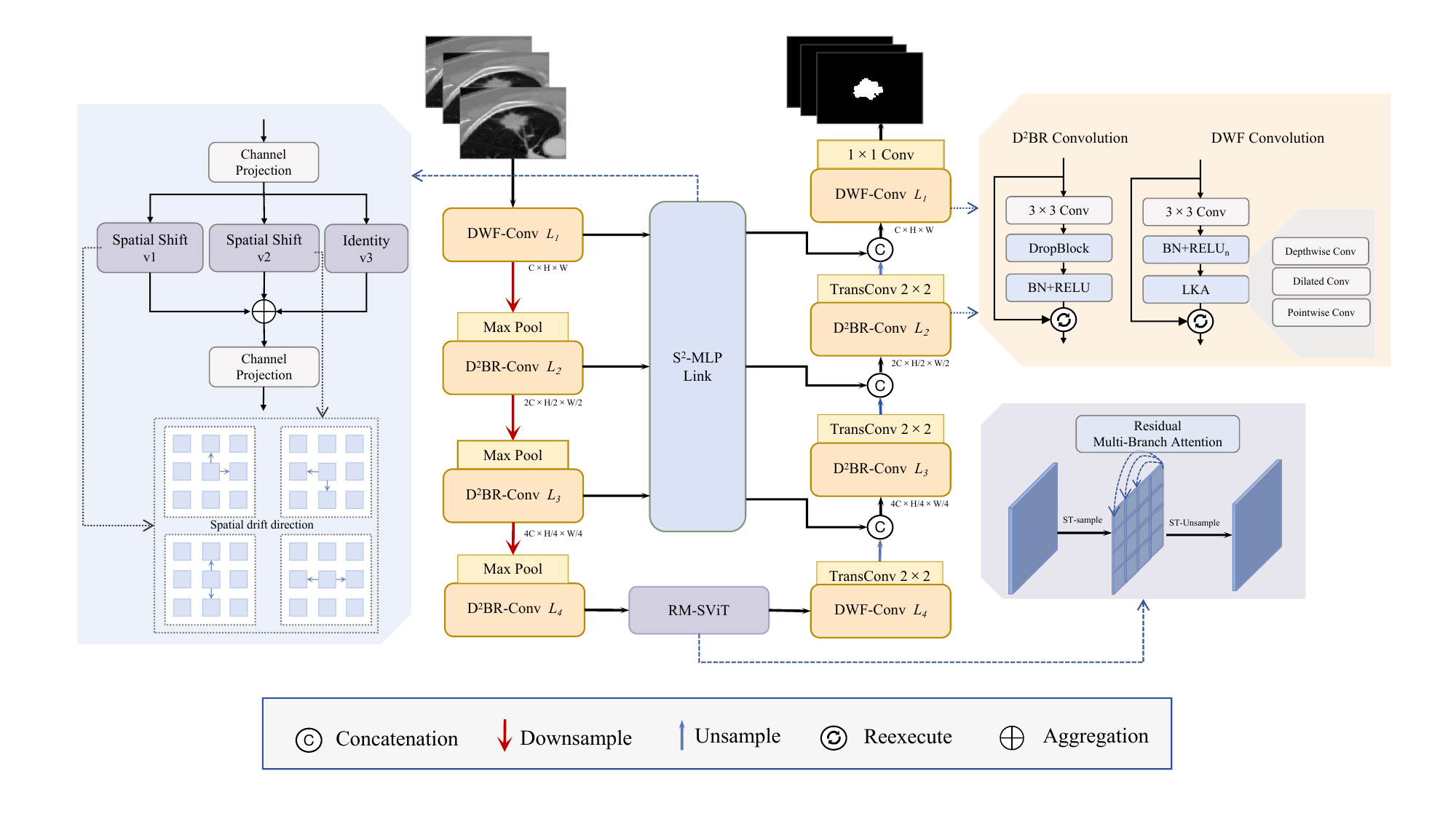}
    \caption{The overall framework of S\textsuperscript{3}TU-Net. The framework is divided into three broad categories of modules, two novel convolutional modules (DWF-Conv/ D\textsuperscript{2}BR-Conv), multi-spatial dimensional connectors (S\textsuperscript{2}-MLP Link), and residual connection-based superpixel vision transformer (RM-SViT).}
    \label{fig:2.1}
\end{figure*}
Lung cancer \cite{xin1,xin2,xin3,lung2,lung3} is one of the most prevalent and fatal cancers globally, with lung adenocarcinoma as the predominant subtype, representing over 50\% of cases \cite{r00}. According to the WHO classification \cite{r0}, lung adenocarcinoma is categorized into distinct stages—atypical adenomatous hyperplasia (AAH), adenocarcinoma in situ (AIS), minimally invasive adenocarcinoma (MIA), and invasive adenocarcinoma (IAC)—each displaying unique imaging characteristics on computed tomography (CT) scans \cite{lung4}. In the early stages, lesions typically appear as ground-glass opacities with clear, regular borders, making detection challenging. In contrast, advanced stages reveal solid, irregular masses with spiculated or lobulated shapes \cite{lung5}.

The manual diagnosis of CT images often risks missing subtle details, making accurate staging difficult. Recent advancements in machine learning and deep learning, including the exploration of diffusion models \cite{xin4,shen2023advancing, shen2024boosting}, have introduced automated classification techniques for lung nodules \cite{lung7}. However, unsegmented CT data introduces excessive computational demands and redundant information, impairing the effectiveness of classification algorithms and decreasing model accuracy \cite{lung6}. Consequently, precise nodule segmentation has become essential, as it underpins accurate classification and provides clinicians with reliable diagnostic insights.

Recent years have seen rapid progress in pulmonary nodule segmentation techniques, shifting from traditional methods \cite{hu2023bag,old2} to deep learning approaches \cite{old6,old7, hu2023robust}. Traditional unsupervised methods, including morphological fuzzy mathematics, threshold segmentation, and fuzzy clustering \cite{old3,old4}, are computationally efficient yet often lack segmentation accuracy. In contrast, deep learning-based methods, such as the widely adopted U-Net \cite{r1,unet2}, leverage a symmetrical encoder-decoder structure with skip connections to combine shallow encoder features with deep decoder features for enhanced segmentation accuracy. Building on this, models like CDP-ResNet \cite{r3} employ a dual-path residual network for multi-view feature extraction and edge-based voxel sampling to capture small nodules. Dense-UNet \cite{r4} mitigates class imbalance issues by using dense connections and an innovative loss function to prevent overfitting and gradient vanishing. The CRF-3D-UNet \cite{r5} further enhances segmentation by integrating conditional random fields with 3D-UNet for spatial and contextual data fusion.

While these convolutional architectures are adept at local feature extraction, they often fall short in capturing long-range dependencies due to limited receptive fields \cite{old8}. The Transformer \cite{r6, yan2024enhancing}, with its self-attention mechanism, provides a solution for long-distance dependency modeling and has shown promise in visual tasks. Vision Transformer (ViT) \cite{r7,r8} exemplifies this approach by effectively modeling global dependencies, though it struggles with fine-grained details and requires significant computational resources and data. Thus, hybrid architectures combining the strengths of convolutional networks and Transformers offer an optimal path forward, where UNet’s efficiency in local feature extraction complements ViT’s global comprehension.

To address these limitations, we propose S\textsuperscript{3}TU-Net, a CNN-Transformer hybrid that integrates multi-space and multi-view fields. S\textsuperscript{3}TU-Net leverages the structured convolutional advantages of CNNs and the global semantic representation capabilities of superpixel-based transformers. By incorporating multi-dimensional spatial connectors for efficient feature fusion, our model enhances feature extraction, fusion, and global context modeling, ultimately improving segmentation performance.
\begin{itemize} \item[\textcolor{black}{$\bullet$}] To overcome U-Net's limitations in handling multi-depth features, we introduce the DWF-Conv block, employing depth-weighted and deep kernel convolutions to improve feature extraction and restoration in the initial encoder-decoder stages. To mitigate overfitting, we propose the D\textsuperscript{2}BR-Conv block, which integrates DropBlock regularization with dual convolutions to reinforce robust feature learning and enhance generalization.

\item[\textcolor{black}{$\bullet$}] Addressing U-Net's challenges in feature fusion, we incorporate multi-dimensional spatial connectors into the skip connections. Specifically, the S\textsuperscript{2}-MLP Link module combines multi-directional spatial shifting and distributed attention mechanisms to integrate features from varied semantic levels, thereby enhancing fusion performance.

\item[\textcolor{black}{$\bullet$}] To improve U-Net's contextual understanding, we propose the RM-SViT module, which fuses global and local features using a multi-branch attention mechanism and superpixel visual transformers. Additionally, residual connections enhance model stability and indirectly improve computational efficiency. \end{itemize}

Experimental results demonstrate that our model achieves a DSC of 89.04\%, precision of 90.73\%, mIoU of 90.70\%, and sensitivity of 93.70\% on the LIDC-IDRI dataset. Furthermore, validation on the independent EPDB private dataset yields a DSC of 86.40\%. These results confirm that S\textsuperscript{3}TU offers high segmentation performance with strong model stability and generalization.

\section{Methods}\label{sec:rw}

\subsection{Network Architecture}  

Fig.\ref{fig:2.1} illustrates the proposed S\textsuperscript{3}TU-Net with a U-shaped encoder-decoder structure. The symmetric S\textsuperscript{3}TU-Net architecture primarily consists of two structured convolution blocks (DWF-Conv/D\textsuperscript{2}BR-Conv) used in the encoder and decoder, fusion residual connections and a multi-branch attention-based superpixel visual transformer (RM-SViT) between the encoder and decoder, and a multi-dimensional spatial connector (S\textsuperscript{2}-MLP Link) based on multi-directional spatial shifting and distributed attention at the skip connections.

Specifically, the encoder's initial stage employs the structured Depth-Weighted Feature Convolution block (DWF-Conv), which consists of two $3 \times 3$ convolutional layers, each followed by batch normalization, a scalable ReLU activation unit, and an LKA module composed of multiple deep kernel convolutions. The encoder then undergoes three downsampling stages, each comprising a structured D\textsuperscript{2}BR-Conv block and $2 \times 2$ max pooling. The D\textsuperscript{2}BR-Conv includes a $3 \times 3$ convolution, DropBlock regularization, batch normalization, and ReLU activation. The RM-SViT, with internal iterative updates, is applied between the encoder and decoder to further enhance feature representation and context understanding. The decoder begins with DWF-Conv, and each upsampling step includes a $2 \times 2$ transpose convolution that halves the number of feature channels. The feature maps from the corresponding encoder layer are processed through the S\textsuperscript{2}-MLP Link module, after which they are concatenated with the upsampled feature maps along the channel dimension. This concatenated result is then passed through the D\textsuperscript{2}BR-Conv block. Finally, the model's output layer employs a $1 \times 1$ convolution and Sigmoid activation to generate the segmentation map.

\subsection{Structured Convolutional Modules}  
\begin{figure}[t]
    \centering
    \includegraphics[width=0.9\linewidth]{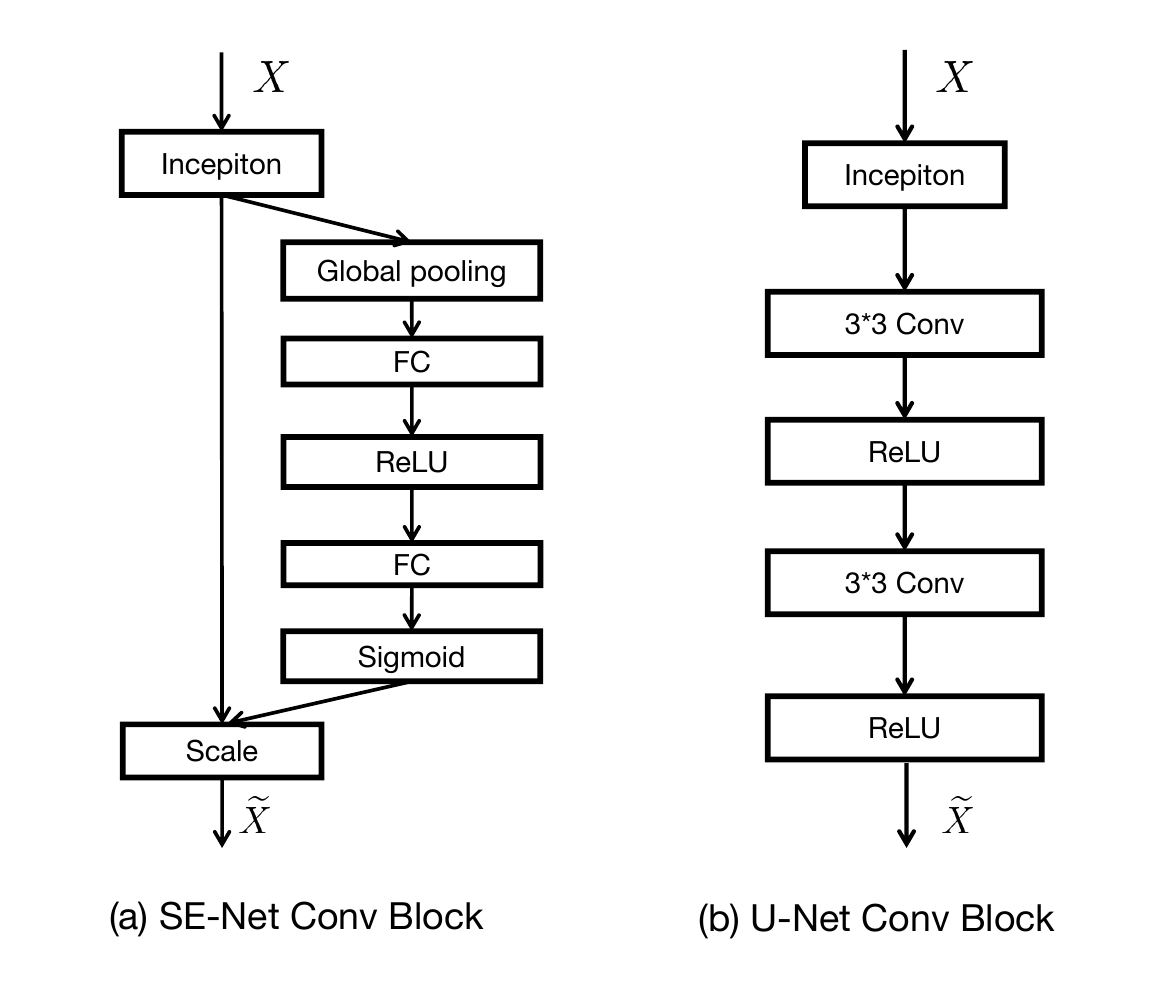}
        \vspace{-0.3cm}
\caption{The architecture of traditional convolutional block. (a) is the convolution module in Squeeze-and-Excitation network. (b) is the traditional convolution module in the UNet network.} 
\label{fig:2.3}  
    \vspace{-0.3cm}
\end{figure}

Traditional U-Net and its various improved versions often suffer from severe overfitting during training. This issue is typically addressed through data augmentation, L2 regularization, or Dropout \cite{r9dropout}, which randomly drops a portion of neurons. In this work, we adopt a more generalized approach—DropBlock \cite{r10dropBlock}. This spatial regularization technique randomly removes contiguous regions from feature maps, forcing the model to make correct predictions even with missing local information. DropBlock controls the size and number of dropped blocks using parameters like block\_size and y.block\_size, effectively preventing overfitting in convolutional networks. Batch normalization is also employed to accelerate the training process, stabilize gradient flow, and prevent issues like vanishing or exploding gradients. Additionally, to enhance feature representation, we incorporate deep large kernel convolutions and dilated convolutions, which capture a wider range of features without increasing computational costs \cite{r11LKA}. We also draw inspiration from the FreeU model \cite{r12FreeU} and Squeeze-and-Excitation Networks (SE-Net) \cite{r13SENet}, which apply feature re-weighting: FreeU improves feature fusion by re-weighting the features between skip connections and backbone feature maps, while SE-Net enhances important features and suppresses irrelevant ones by channel-wise re-weighting after each convolutional layer using global average pooling and fully connected layers (Fig. \ref{fig:2.3}(a)).

\begin{figure}[t]
    \centering
    \includegraphics[width=1\linewidth]{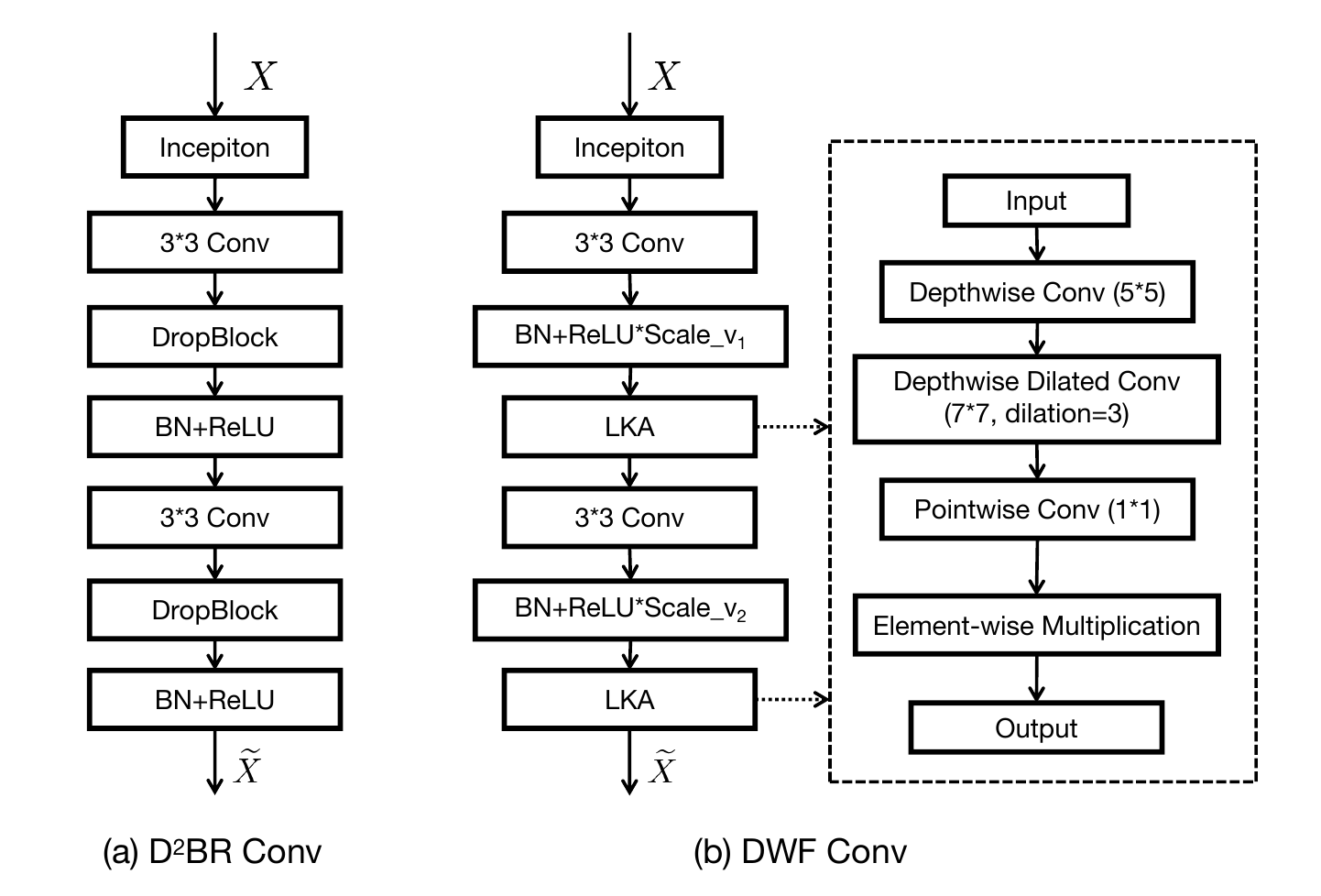}
        \vspace{-0.3cm}
\caption{The architecture of Newly proposed convolutional block. (a) is the convolutional block named D\textsuperscript{2}BR, which is composed of $3 \times 3$ Conv, DropBlock, BN, and ReLU. (b) is a convolutional block named DWF, which is composed of a combination of $3 \times 3$ Conv, BN, LKA module, and ReLU with different scaling weight values. The LKA module contains multiple large kernel convolutions, depth convolutions, and pointwise convolutions that can expand the receptive field.} 
\label{fig:2.4}  
    \vspace{-0.3cm}
\end{figure}

Based on these insights, we designed two structured convolutional blocks: the Deep Weighted Feature Convolution (DWF-Conv) and the Double Drop Convolution (D\textsuperscript{2}BR-Conv). DWF-Conv is used at the beginning stages of both the encoder and decoder. It leverages LKA to focus on a broader range of features and utilizes scalable ReLU to enhance feature expression, aiding in the comprehensive capture of global information and the effective restoration of the overall image structure. D\textsuperscript{2}BR-Conv is employed multiple times in the middle stages of the U-shaped network, utilizing DropBlock regularization to enforce the learning of more robust features. As shown in Fig. \ref{fig:2.4}(a), D\textsuperscript{2}BR-Conv consists of a DropBlock, a Batch Normalization (BN) layer, and a ReLU activation unit following each convolutional layer. As shown in Fig. \ref{fig:2.4}(b), each convolutional layer in the DWF-Conv is immediately followed by BN, a flexible ReLU unit with adjustable feature weighting parameters, and multiple deep large kernel attention (LKA) layers. This approach enhances model performance without additional computational costs by introducing large kernel convolutions and adjustable weight parameters at specific layers. Unlike the original convolutional blocks in U-Net (Fig. \ref{fig:2.3}(b)), these structured convolutional blocks mitigate overfitting while accelerating network convergence.

\subsection{RM-SViT Module}  
\begin{figure}[t]
    \centering
    \includegraphics[width=1\linewidth]{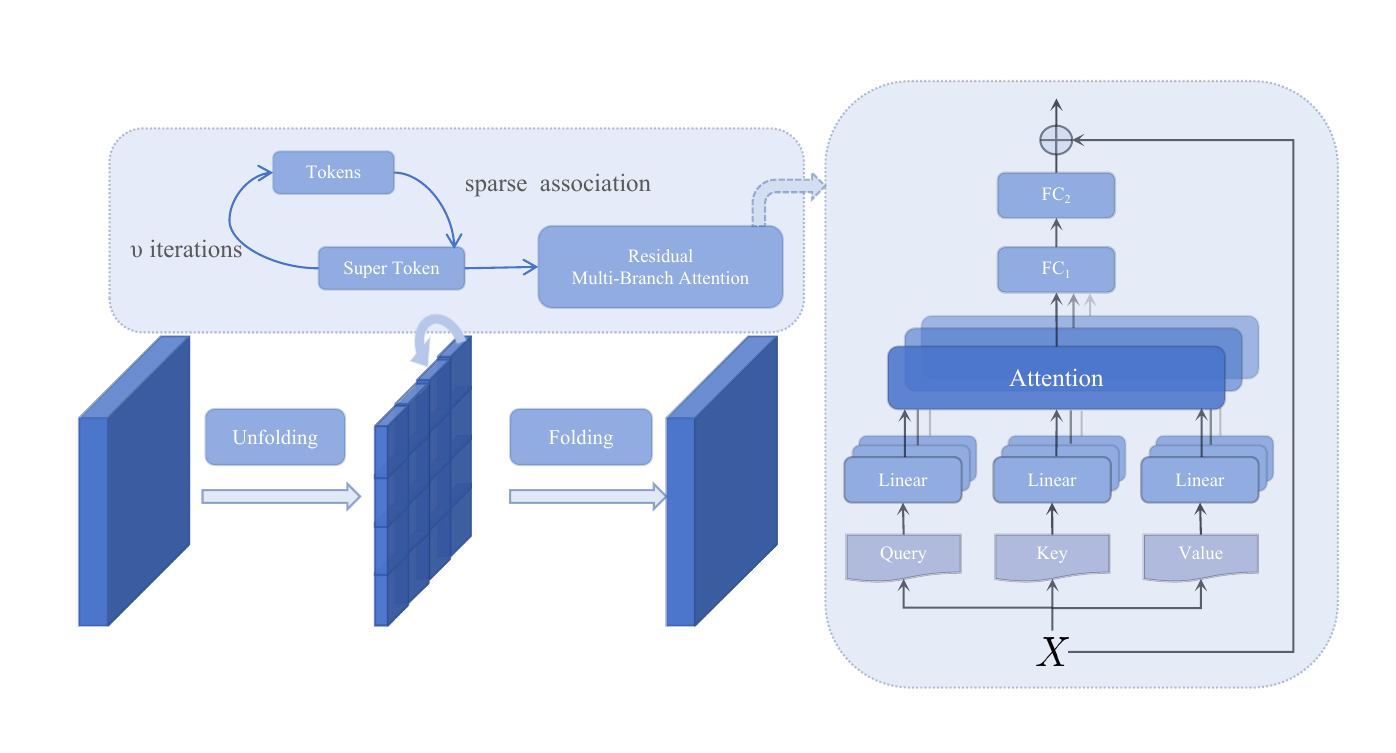}
        \vspace{-0.3cm}
\caption{The architecture of RM-SViT Module. The encoder expands the feature tensor, divides 'Tokens' into 'Super tokens' by sparse association learning, then adjusts the final 'Super Token' by applying multi-branch self-attention based on residual connection after corresponding rounds of iteration, and finally maps the expanded local block back to the original Token space.}
\label{fig:2.6}
    \vspace{-0.3cm}
\end{figure}

To enhance the network's ability to model global context information, we propose the residual and multi-branch attention based superpixel vision transformer (RM-SViT) module, which integrates residual connections and multi-branch attention with superpixel visual transformers. Integrated between the encoder and decoder of the U-shaped network, the RM-SViT module (Fig. \ref{fig:2.6}) iteratively samples visual tokens through sparse relational learning. It then applies residual multi-branch attention (RMBA) on the superpixels, merging the features before mapping them back to the original tokens.
The execution process of the RM-SViT module begins by unfolding the feature tensor \(F_{enc}\) extracted by the encoder into non-overlapping local patches, \( F_{unfold} \), and then dividing them into initial superpixels \( S_0 \). The superpixels are initialized by averaging the features within each grid area. If the grid size is \( h \times w \), the number of superpixels is given by Eq. \textcolor{red}{(1)}:
\begin{equation}
 m = \frac{H}{h} \times \frac{W}{w}.
\end{equation}
\label{eq:1}
This method ensures an even distribution of superpixels across the image, providing a solid starting point for iterative updates. For each iteration \( t \), the association \( Q^t_{ij} \) between feature \( X_i \) and superpixel \( S_j \) is calculated using the following Eq. \textcolor{red}{(2)}:
\begin{equation}
Q^t_{ij} = \text{Softmax}\left(\frac{X_i (S^{t-1})_j^T}{\sqrt{d}}\right).
\end{equation}
\label{eq:2}
where \( d \) is the number of channels \( C \). Subsequently, the super token \( S \) is updated as the weighted sum of tokens, as in Eq. \textcolor{red}{(3)}:
\begin{equation}
S = (\hat{Q}^t)^T X.
\end{equation}
\label{3}
where \( \hat{Q}^t \) is the column-normalized version of \( Q^t \). After several iterations, multi-branch self-attention is applied to adjust the final superpixel \( S \), capturing global context dependencies. In this Eq. \textcolor{red}{(4)}:
\begin{equation}
Q = q(S), \quad K = k(S), \quad V = v(S).
\end{equation}
\label{4}
Scaled dot-product attention is used to compute the attention weights, normalized by Softmax, and then a weighted sum of values \( V \)is performed along the last dimension, as in Eq. \textcolor{red}{(5)}:
\begin{equation}
\text{Attn}(S) = \text{Softmax}\left(\frac{Q K^T}{\sqrt{d}}\right) V .
\end{equation}
\label{5}
The result of the weighted sum is then projected through a convolutional layer and added to the residual connection. The output is finally obtained by combining the adjusted features with the residual connection,as in Eq. \textcolor{red}{(6)}:
\begin{equation}
\text{output} = \text{LayerNorm}(\text{Conv2d}(\text{Attn}(S)) +\text{residual}).
\end{equation}
\label{6}
\subsection{S\textsuperscript{2}-MLP Link Module}  
The skip connections enhance the transmission of information between multi-scale feature maps, thereby improving the feature-capturing ability of deep networks \cite{r15Ruan}. To further gain spatial perception across different dimensions and understand complex positional relationships, we introduce a multi-dimensional spatial connector at the skip connections, namely the spatial-shift mlp (S\textsuperscript{2}-MLP Link) module. Spatial shift mlp methods \cite{r16Yu}\cite{r17Tols},combine the inductive bias advantages of MLPs with spatial shifting to enable patch communication, thus achieving high recognition accuracy.
 
\begin{figure}[t]
    \centering
    \includegraphics[width=1\linewidth]{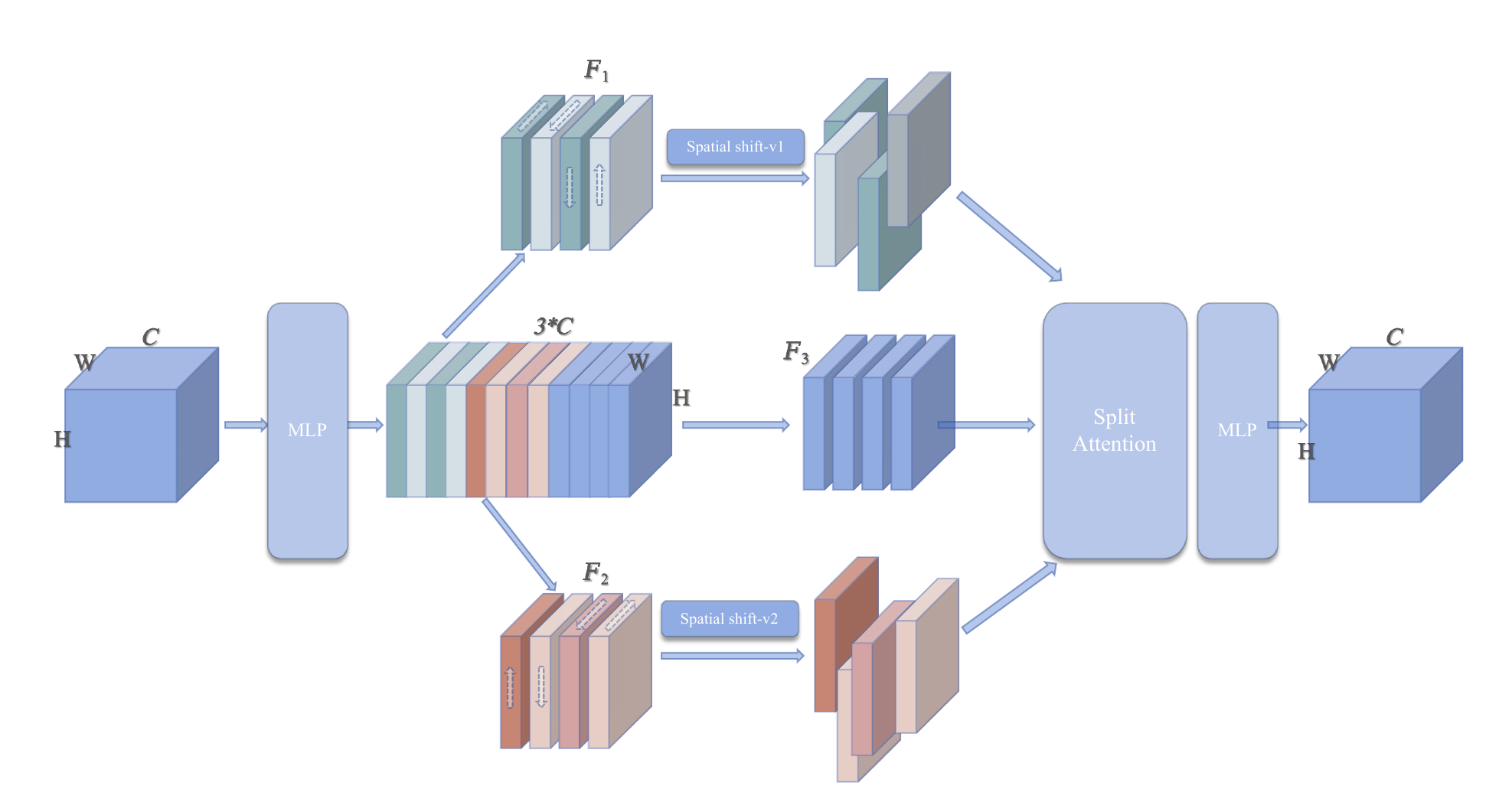}
\caption{The architecture of S\textsuperscript{2}-MLP Link Module. Firstly, MLP is used to expand the channel c of the feature map into $3 \times c$ and divide it into three parts (\(F_{1}\),\(F_{2}\),\(F_{3}\)) along the channel dimension. \(F_{1}\) and \(F_{2}\) are spatially shifted according to different directions, and \(F_{3}\) remains unchanged. Then, Split Attention is used for weighting calculation, and finally MLP is used for recovery.}
\label{fig:2.7}
\end{figure}

As a multi-dimensional spatial connector, the S\textsuperscript{2}-MLP Link Module, as shown in Fig. \ref{fig:2.7}, consists of an MLP as the patch embedding layer, a spatial shifting module, and a SplitAttention module. First, the MLP$_{1}$ expands the feature map's channel dimension \([b,c,h,w]\) to three times its original size, splitting it into three parts (\(F_{1}\),\(F_{2}\),\(F_{3}\)). Spatial shifts are applied to \(F_{1}\) and \(F_{2}\), while \(F_{3}\) remains unchanged. The parts are then stacked into a tensor \([b,2,h*w,c]\). The Split Attention module calculates and applies attention weights to the stacked features. Finally, the MLP$_{2}$ restores the weighted features, producing the output feature map.

\subsubsection{MLP Layer}
In Vision Transformers (ViT) , Patch Embedding divides the input image into small patches, converting each into a high-dimensional embedding vector. The MLP plays a crucial role in both Patch Embedding and the final classification head \cite{r18MLP}. The initial MLP rearranges the dimensions of the input feature map and expands each pixel block into a high-dimensional vector with three times the number of channels.The subsequent MLP linearly transforms the attention-enhanced feature map, restoring the original channel count.

\subsubsection{Spatial Shift Block}
In the S\textsuperscript{2}-MLP proposed by Yu et al. \cite{r16Yu}, the spatial shift concept divides the \( c \) channels into four parts, each shifted in different directions—up, down, left, and right. In the S\textsuperscript{2}-MLP Link, the spatial shift module similarly shifts different parts of the channels in various directions, enhancing the capture of spatial contextual information and improving the model's performance and generalization in complex visual tasks.
In Eq. \textcolor{red}{(7)} and \textcolor{red}{(8)}, the first 1/4 of the channels are shifted left (up) by one row (column), filled with the next row's (column's) values; the next 1/4 to 1/2 of the channels are shifted right (down) by one row (column), filled with the previous row's (column's) values; the 1/2 to 3/4 of the channels are shifted up (left) by one column (row), filled with the next column's (row's) values; and the final 3/4 of the channels are shifted down (right) by one column, filled with the previous column's values. The formulas below illustrate the spatial shift operations for the different channel groups.
\begin{equation}
\text{SS1}(x) = 
\begin{cases}
x[:, :w - 1, :, :\frac{c}{4}] & \text{if } x[:, 1:, :, :\frac{c}{4}] \\
x[:, 1:, :, \frac{c}{4}:\frac{c}{2}] & \text{if } x[:, :w - 1, :, \frac{c}{4}:\frac{c}{2}] \\
x[:, :, :h - 1, \frac{c}{2}:\frac{3c}{4}] & \text{if } x[:, :, 1:, \frac{c}{2}:\frac{3c}{4}] \\
x[:, :, 1:, \frac{3c}{4}:] & \text{if } x[:, :, :h - 1, \frac{3c}{4}:]
\end{cases}
\end{equation}
\label{7}
\begin{equation}
\text{SS2}(x) = 
\begin{cases}
x[:, :, :h - 1, :\frac{c}{4}] & \text{if } x[:, :, 1:, :\frac{c}{4}] \\
x[:, :, 1:, \frac{c}{4}:\frac{c}{2}] & \text{if } x[:, :, :h - 1, \frac{c}{4}:\frac{c}{2}] \\
x[:, :w - 1, :, \frac{c}{2}:\frac{3c}{4}] & \text{if } x[:, 1:, :, \frac{c}{2}:\frac{3c}{4}] \\
x[:, 1:, :, \frac{3c}{4}:] & \text{if } x[:, :w - 1, :, \frac{3c}{4}:]
\end{cases}
\end{equation}
\label{8}
\subsubsection{Split Attention}
Split Attention is derived from the ResNest model proposed, where feature maps are finely divided, transformed, fused within groups, and then weighted and summed using attention mechanisms. A residual connection then produces the final feature map with diverse information. This paper adopts the core idea: leveraging multi-head attention and global context to perform weighted fusion on input feature maps, enhancing the diversity and accuracy of feature representation. Additionally, feature reshaping and normalization steps ensure the model's stability in complex tasks.The main processing steps of this module are as follows:

\begin{table}[t]
\caption{The information of LIDC-IDRI and EPDB datasets}
\label{table:3.1}
\begin{center}
\renewcommand{\arraystretch}{1.5} 
\begin{tabular}{ >{\centering\arraybackslash}m{1.5cm} 
>{\centering\arraybackslash}m{3cm} 
>{\centering\arraybackslash}m{3cm} 
>{\centering\arraybackslash}m{3cm}}
 \Xhline{1.1pt}  
 \textbf{Dataset} & \textbf{LIDC-IDRI}  &  \textbf{EPDB} \\  
 \Xhline{1.1pt}  
 Source & National Cancer Institute & Shanghai Electric Power Hospital \\
 Number & 6474 & 1198 \\ 
 Partition & Train:Test=9:1 & Valid \\ 
 Resize & 128*128  & 128*128 \\ 
 \Xhline{1.2pt}  
\end{tabular}
\end{center}
\end{table}

\begin{table}[t]
\caption{System Configuration Information}
\label{table:3.2}
\begin{center}
\renewcommand{\arraystretch}{1.5} 
\begin{tabular}{ >{\centering\arraybackslash}m{4cm} 
>{\centering\arraybackslash}m{4cm}}
 \Xhline{1.1pt}  
 \textbf{Configuration Name} & \textbf{Details} \\  
 \Xhline{1.1pt}  
 Operating System & Linux ai 5.15.0-86-generic \\ 
 CPU Model & Intel(R) Core(TM) i7-10700K CPU @ 3.80GHz \\ 
 GPU Model & NVIDIA GeForce RTX 3080 \\ 
 Memory Information & 20GB \\ 
 Python Version & 3.8.10 \\ 
 Cuda Version & 12.2 \\ 
 PyTorch Version & 1.13.1 \\ 
 \Xhline{1.1pt}  
\end{tabular}
\end{center}
\end{table}

First, the input tensor \( x_{\text{all}} \) is reshaped to the form \((b, k, h, w, c)\), where \( b \) is the batch size, \( k \) is the number of attention heads, \( h \) and \( w \) are height and width, and \( c \) is the number of channels. The tensor is further reshaped to \((b, k, n, c)\), where \( n = h \times w \), to facilitate matrix operations. 
According to Eq. \textcolor{red}{(9)}, the input tensor is summed over the spatial and head dimensions and averaged to obtain the intermediate representation \( a \):
\begin{equation}
a = \frac{1}{k} \sum_{i=1}^{k} \left( \frac{1}{n} \sum_{j=1}^{n} x_{\text{all}, i, j} \right) = \frac{1}{kn} \sum_{i=1}^{k} \sum_{j=1}^{n} x_{\text{all}, i, j}.
\end{equation}
\label{9}
The intermediate representation \( a \) is then passed through two layers of MLP and a GELU activation function to compute a higher-dimensional representation \( \hat{a} \). First, as in Eq. \textcolor{red}{(10)}, \( a \) is reduced to \( c/2 \) dimensions through the first MLP, activated by GELU, and then expanded to \( kc \) dimensions by the second MLP:
\begin{equation}
\hat{a} = \text{MLP}_2(\text{GELU}(\text{MLP}_1(a))) \in \mathbb{R}^{(b, kc)}.
\end{equation}
\label{10}
Next, as in Eq. \textcolor{red}{(11)}, \( \hat{a} \) is reshaped to \((b, k, c)\) and normalized using the Softmax function to obtain the attention weights \( \bar{a} \):
\begin{equation}
\bar{a} = \text{Softmax}(\hat{a}) \in \mathbb{R}^{(b, k, c)}.
\end{equation}
\label{11}
Once the attention weights \( \bar{a} \) are obtained, they are expanded by one dimension to be element-wise multiplied with the input tensor, producing the attention-weighted matrix. The weighted feature maps are then reshaped to generate the final output feature map \( \hat{X} \) as in Eq. \textcolor{red}{(12) (13) (14)}:  
\begin{equation}
\text{attention} = \bar{a} \in \mathbb{R}^{(b, k, 1, c)},
\end{equation}
\label{12}
\begin{equation}
\text{out}_{ijkc} = x_{\text{all}, ijkc} \cdot \text{attention}_{ijkc},
\end{equation}
\label{13}
\begin{equation}
\hat{X} = \sum_{k=1}^{K} \text{out}_{ijkc} \rightarrow \hat{X} \in \mathbb{R}^{(b, h, w, c)}.
\end{equation}
\label{14}

\section{Experiment and Analysis}\label{sec:exp}  
\begin{figure*}[t]
    \centering
\includegraphics[width=0.9\linewidth]{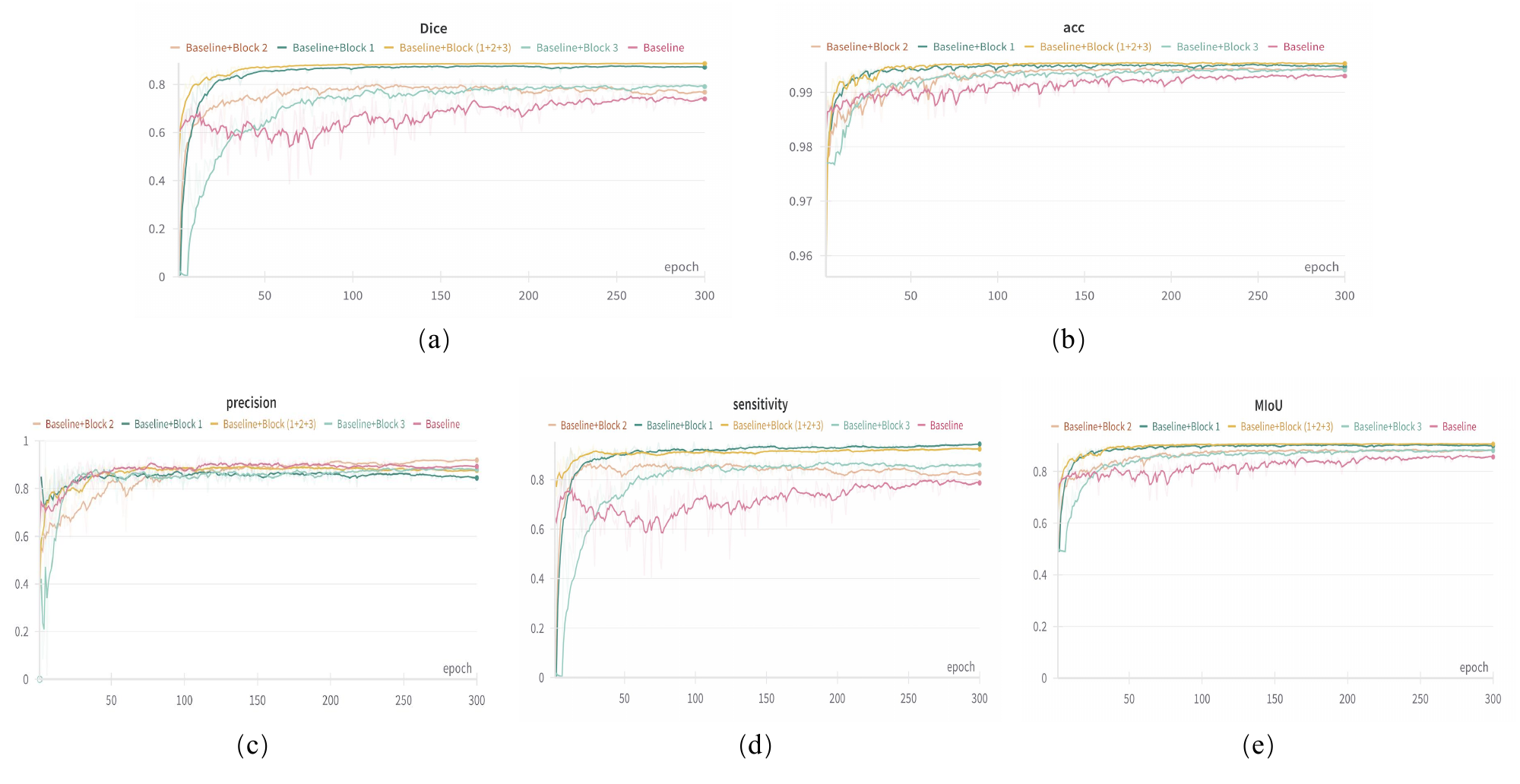}
    \vspace{-0.3cm}
\caption{Comparison results of performance. Various performance comparison results on LIDC-IDRI dataset (Baseline/ +Block$_{1}$/ +Block$_{2}$/ +Block$_{3}$), The performance metrics from graph (a) to graph (e) are Dice, Accuracy, Precision, Sensitivity, and MIoU.}
\vspace{-.2cm}
\label{fig:LIDC performance compare}
\end{figure*}

This section covers three parts: the two datasets used in the experiments, the evaluation metrics employed, and the implementation details.
\subsection{Datasets}
We evaluated S\textsuperscript{3}TU-Net on two datasets: the public LIDC-IDRI \cite{r20LIDC} and the private EPDB from Shanghai Electric Power Hospital. LIDC-IDRI is a public dataset from the Lung Image Database Consortium and Image Database Resource Initiative. We prioritized nodules annotated by at least two radiologists, extracting valid nodule data from XML files, resulting in 1,303 nodules across 6,474 CT slices. The EPDB dataset includes samples from 481 lung adenocarcinoma patients diagnosed between September 2016 and October 2023, with annotations verified by three experienced radiologists. To validate segmentation performance, we selected 1,198 CT slices containing four types of lung adenocarcinoma (AAH/MIA/IAC/AIS). Table. \ref{table:3.1} provides details. Preprocessing steps, including lung parenchyma segmentation, ROI extraction, and image enhancement, were applied. The final input was 128×128 patches centered on the lung nodules.

\begin{table*}[t]
\vspace{0.3cm}
\caption{Performance Comparison on LIDC-IDRI Dataset (Max)}
\label{table:performance_lidc_max}
\begin{center}
\renewcommand{\arraystretch}{1.5} 
\begin{tabular}{>{\centering\arraybackslash}m{3cm} 
>{\centering\arraybackslash}m{2.5cm} 
>{\centering\arraybackslash}m{2.5cm} 
>{\centering\arraybackslash}m{2.5cm} 
>{\centering\arraybackslash}m{2.5cm} 
>{\centering\arraybackslash}m{2.5cm}}
 \Xhline{1.1pt}  
 \textbf{Methods} & \textbf{DSC \%} & \textbf{Acc \%} & \textbf{mIoU \%} & \textbf{Pre \%} & \textbf{Sen \%} \\  
 \Xhline{1.1pt}  
 Baseline & 77.44 & 99.33 & 86.41 & 93.14 & 86.94 \\ 
 + Block1 & 86.85 & 99.52 & 90.31 & 87.40 & \textbf{95.35} \\ 
 + Block2 & 82.39 & 99.41 & 88.53 & \textbf{93.25} & 88.56 \\ 
 + Block3 & 82.35 & 99.40 & 88.64 & 90.34 & 90.53 \\ 
 + ALL (S\textsuperscript{3}TU-Net) & \textbf{89.04} & \textbf{99.53} & \textbf{90.70} & 90.73 & 93.70 \\ 
 \Xhline{1.1pt}  
\end{tabular}
\end{center}
\vspace{-0.3cm}
\end{table*}

\begin{table*}[t]
\vspace{0.3cm}
\caption{Performance Metrics on EPDB Dataset (Average)}
\label{table:performance_epdb_avg}
\begin{center}
\renewcommand{\arraystretch}{1.5} 
\begin{tabular}{>{\centering\arraybackslash}m{3cm} 
>{\centering\arraybackslash}m{2.5cm} 
>{\centering\arraybackslash}m{2.5cm} 
>{\centering\arraybackslash}m{2.5cm} 
>{\centering\arraybackslash}m{2.5cm} 
>{\centering\arraybackslash}m{2.5cm}}
 \Xhline{1.1pt}  
 \textbf{Methods} & \textbf{DSC \%} & \textbf{Acc \%} & \textbf{AUC \%} & \textbf{Pre \%} & \textbf{Sen \%} \\  
 \Xhline{1.1pt}  
 + Block1 & 85.01 & 98.38 & 89.04 & 94.54 & 78.36 \\ 
 + Block2 & 81.55 & 98.14 & 86.30 & 95.49 & 72.85 \\ 
 + Block3 & 83.13 & 98.44 & 86.94 & \textbf{96.38} & 74.04 \\ 
 + ALL (S\textsuperscript{3}TU-Net) & \textbf{86.40} & \textbf{98.53} & \textbf{90.00} & 94.79 & \textbf{80.27} \\ 
 \Xhline{1.1pt}  
\end{tabular}
\end{center}
\vspace{-0.3cm}
\end{table*}

\subsection{Evaluation Metrics}  
This study assesses segmentation performance using common metrics in medical image segmentation: Dice Similarity Coefficient (DSC), Accuracy (Acc), Sensitivity (Sen), Precision (Pre), and supplementary metrics like IoU or AUC. DSC measures the similarity between the predicted mask and the ground truth, Acc measures the percentage of correctly classified pixels, Sen evaluates the model's ability to detect true positives, and Pre indicates the proportion of true positives among predicted positives. \( A \) and \( B \) represent the prediction and ground truth; \( N \) is the number of pixels in the 3D patch; \( p_i \) is the predicted probability for pixel \( i \); \( g_i \) is the ground truth label for pixel \( i \). The Eq. \textcolor{red}{(15) (16) (17)} are as follows:
\begin{equation}
\text{DSC} = \frac{2 |A \cap B|}{|A| + |B|} = \frac{2 \sum_{i=1}^{N} p_i g_i}{\sum_{i=1}^{N} p_i^2 + \sum_{i=1}^{N} g_i^2},
\end{equation}
\label{15}
\begin{equation}
\text{Sensitivity} = \frac{|A \cap B|}{|B|} = \frac{\sum_{i=1}^{N} p_i g_i}{\sum_{i=1}^{N} g_i},
\end{equation}
\label{16}
\begin{equation}
\text{Precision} = \frac{|A \cap B|}{|A|} = \frac{\sum_{i=1}^{N} p_i \cdot g_i}{\sum_{i=1}^{N} p_i}.
\end{equation}
\label{17}
\subsection{Implementation Details} 
The dataset was divided into training, testing, and independent validation sets to more accurately evaluate the model's segmentation performance and generalization. The LIDC dataset was split with a nearly 9:1 ratio, with 5,717 images for training and 757 for testing. The EPDB dataset's 1,198 images were used as an independent validation set to objectively assess the model's generalization and prepare for subsequent lung adenocarcinoma nodule classification. The S\textsuperscript{3}TU-Net model was trained using the Adam optimizer with a combined binary cross-entropy and Dice loss function. The initial learning rate was set at 0.001, dynamically adjusted using a scheduler, with one warm-up epoch. The batch size was 16, with 300 epochs, and the DropBlock size was set to 7. The implementation was based on the PyTorch framework, with the hardware and software configurations detailed in Table. \ref{table:3.2}.

\section{Experimental Results}\label{sec:method}
\subsection{Ablation Study}

\begin{figure*}[t]
    \centering
    \includegraphics[width=0.9\linewidth]{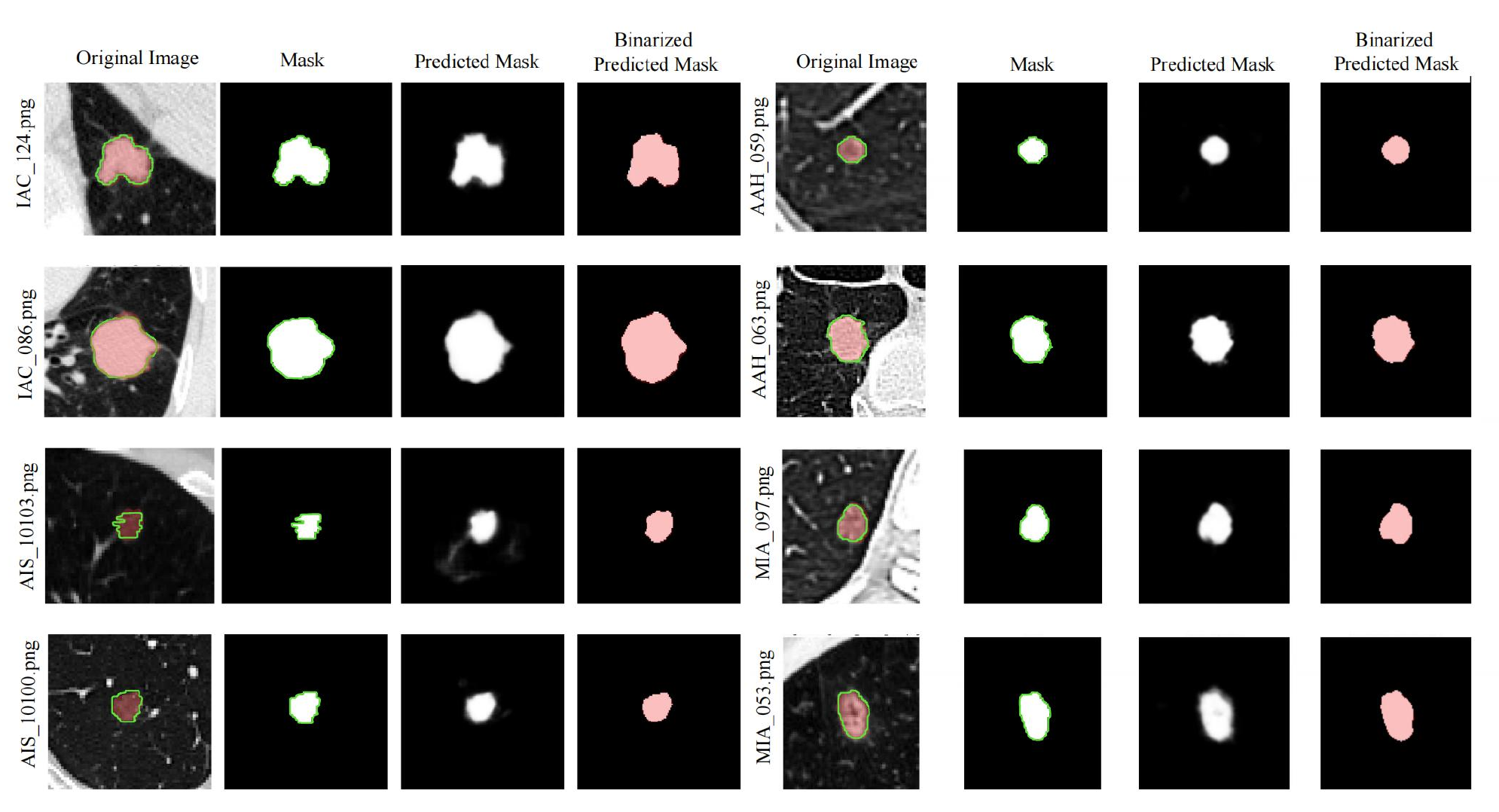}
\caption{The Example Images From the EPDB Dataset. Randomly shown are the original images, annotated masks, segmentation results, and binarized segmentation results of stage IV lung adenocarcinoma (AAH /MIA /IAC /AIS).}
\vspace{-.2cm}
\label{fig:example images from the EPDB dataset}
\end{figure*}

To demonstrate that each component of the proposed S\textsuperscript{3}TU-Net enhances lung nodule segmentation performance, we conducted ablation experiments on the LIDC-IDRI dataset and validated the results on the EPDB dataset. Using the UNet model as the baseline, we evaluated the segmentation performance of Baseline+Block$_{1}$ (structured convolution), Baseline+Block$_{2}$ (RM-SViT), and Baseline+Block$_{3}$ (S\textsuperscript{2}-MLP Link) on both datasets, as shown in Table. \ref{table:performance_lidc_max} and Table. \ref{table:performance_epdb_avg}. Fig. \ref{fig:LIDC performance compare} present the corresponding performance comparison on the LIDC-IDRI dataset. Since the segmentation task is a precursor to lung adenocarcinoma classification, which involves various irregular nodule shapes, the EPDB dataset was used to test the model's generalization ability. Fig. \ref{fig:example images from the EPDB dataset} shows the segmentation results of the four types of lung adenocarcinoma (AAH, MIA, IAC, AIS) on sample nodules from the EPDB dataset.
\subsubsection{Baseline+Block$_{1}$}
When replacing traditional convolutions with two types of structured convolutions, performance improvements were most notable. The DSC reached 86.85\%, mIoU was 90.31\%, and sensitivity peaked at 95.35\%. Compared with the Baseline model, the DSC and sensitivity are increased by 9.41\%. This demonstrates the effectiveness of constructing the backbone using novel structured convolutional blocks with distinct functions.
\subsubsection{Baseline+Block$_{2}$}
\begin{figure*}[t]
    \centering
    \includegraphics[width=0.9\linewidth]{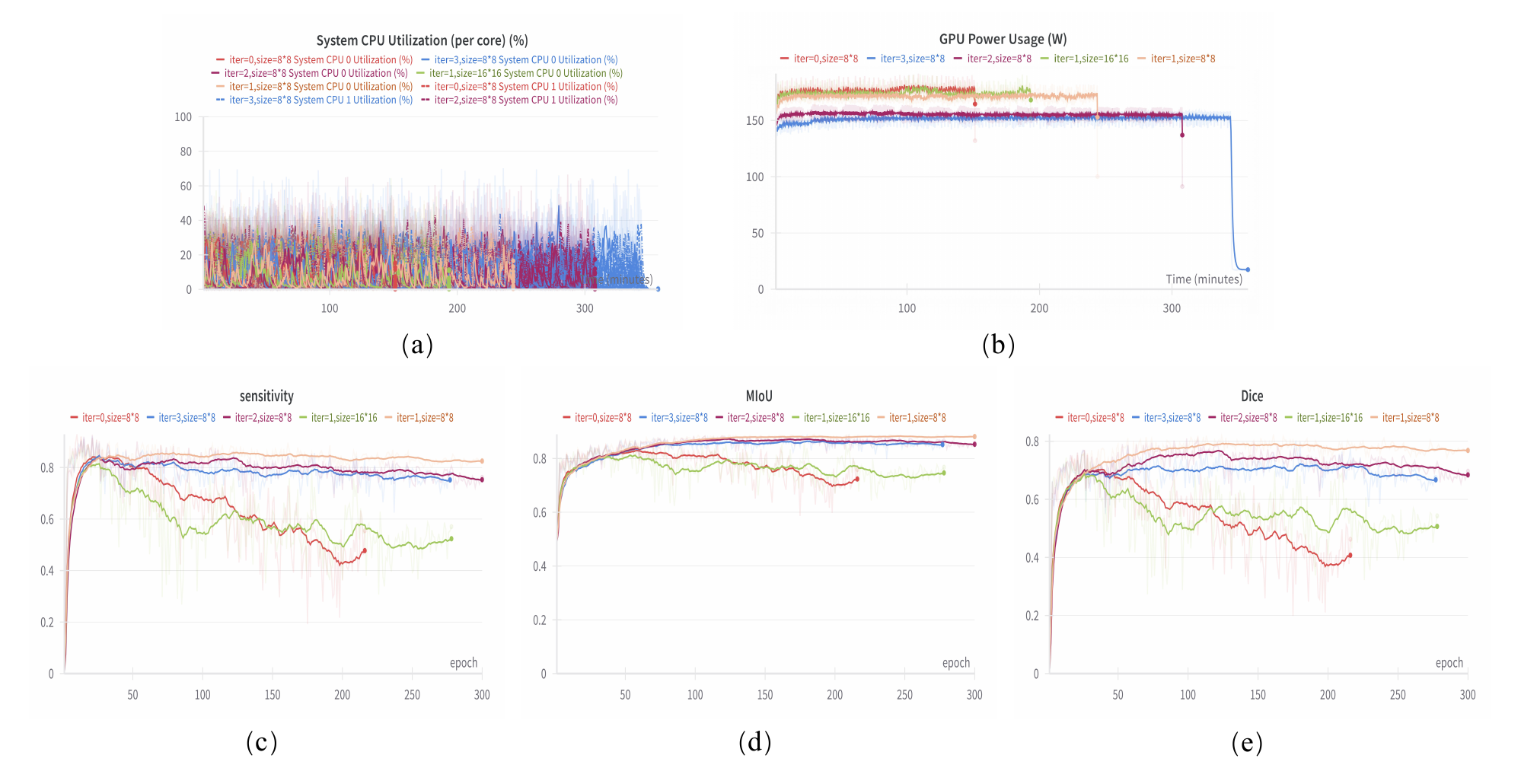}
    \vspace{-0.3cm}
\caption{Performance comparison on Block$_{2}$. iter=0/1/2/3 and size=$8 \times 8$/ $16 \times 16$.(a) System CPU utilization. (b) GPU power consumption. (c) Sensitivity. (d) MIoU. (e) Dice. (left to right, top to bottom)}
\vspace{-.3cm}
\label{fig:partall}
\end{figure*}

RM-SViT was developed to address limitations in capturing long-range dependencies and global context during local feature extraction. Experiments showed that the size of the super token coverage (stoken\_size) and the number of iteration updates strongly affect segmentation performance and computational complexity. A smaller grid size (8×8) captures local features more precisely but increases computational cost, while a larger grid size (16×16) focuses on global features and reduces complexity. Higher iteration counts capture more complex relationships between image patches but significantly increase processing time. Without iterations, performance drops due to potential misalignment of super tokens across different semantic regions. Fig.\ref{fig:partall} shows the performance comparison results, including the system CPU utilization, GPU power consumption, sensitivity, MIoU, and Dice. The final conclusion is that a single iteration with a smaller grid size is best suited for processing 128×128 input images.Excessive or insufficient iterations affect Dice accuracy, wasting computation time despite low GPU energy consumption. An 8×8 grid size outperforms 16×16 in Dice, mIoU, and sensitivity under similar GPU and CPU consumption. Compared to the baseline, the setup with one iteration and an 8×8 grid size improves DSC by 4.95\%, mIoU by 2.12\%, and sensitivity by 1.64\%.
\subsubsection{Baseline+Block$_{3}$}
The S\textsuperscript{2}-MLP Link optimizes skip connections by leveraging spatial information to enhance feature fusion and aid gradient flow. Compared to the baseline, with only a slight increase in parameters, it achieves a DSC of 82.35\%, an accuracy of 99.40\%, mIoU of 88.64\%, and sensitivity of 90.53\%. This further demonstrates that MLP-based visual architectures combined with spatial shifts can achieve higher performance with less inductive bias.
\begin{figure*}[ht]
    \centering
    \includegraphics[width=0.8\linewidth]{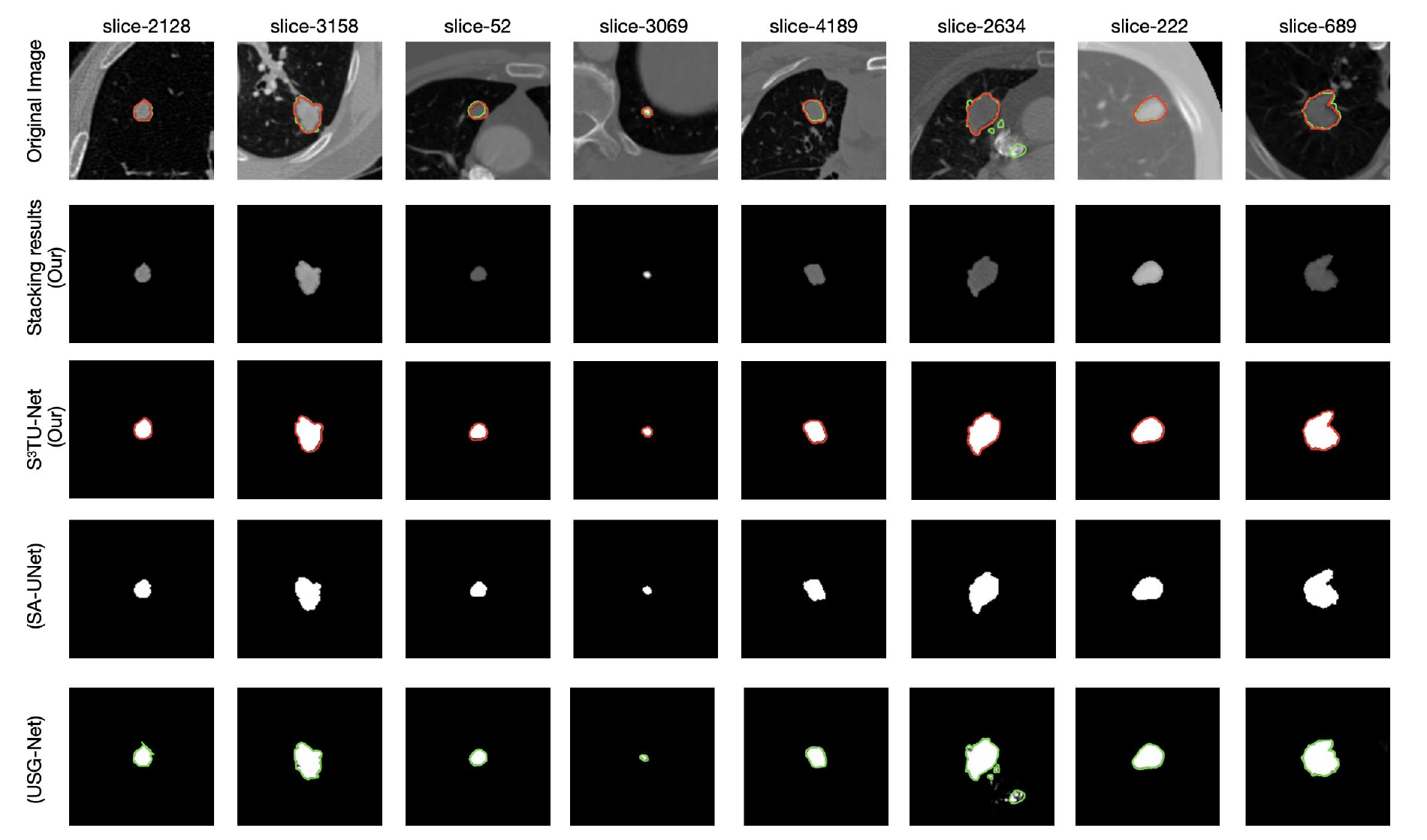}
        \vspace{-0.3cm}
\caption{Qualitative comparison (data were randomly extracted from LIDC-IDRI).(1) Original image. (2/3) Segmentation results and display results of S\textsuperscript{3}TU-Net. (4/5) The segmentation results of the comparison model.}
\vspace{-.3cm}
\label{fig:datu}
\end{figure*}
\begin{table*}[t]
\vspace{0.3cm}
\caption{Performance Comparison of Different Methods}
\label{table:performance_comparison_methods}
\begin{center}
\renewcommand{\arraystretch}{1.5} 
\begin{tabular}{>{\centering\arraybackslash}m{3cm} 
>{\centering\arraybackslash}m{2cm} 
>{\centering\arraybackslash}m{2cm} 
>{\centering\arraybackslash}m{2cm} 
>{\centering\arraybackslash}m{2cm} 
>{\centering\arraybackslash}m{2cm}}
 \Xhline{1.1pt}  
 \textbf{Methods} & \textbf{DSC \%} & \textbf{Acc \%} & \textbf{mIoU \%} & \textbf{Pre \%} & \textbf{Sen \%} \\  
 \Xhline{1.1pt}  
 Y-Net\cite{r22YNet} & 79.53 & 92.39 & 83.71 & 83.77 & 78.54 \\ 
 UNet++\cite{r23UNet++} & 82.27 & 96.28 & 85.37 & 82.93 & 84.98 \\ 
 R2U-Net\cite{r26R2UNet} & 82.10 & 98.11 & 76.18 & 84.00 & 86.33 \\ 
 Attention-UNet\cite{r24AttentionNet} & 80.21 & 98.78 & 82.84 & 83.78 & 86.91 \\ 
 UNet3++\cite{r28UNet3++} & 81.34 & \textcolor{red}{99.80} & 83.64 & 81.14 & 84.50 \\ 
 SA-UNet\cite{r25SAUNet} & \textcolor{green}{84.35} & \textcolor{green}{99.24} & \textcolor{blue}{88.70} & \textcolor{blue}{88.20} & \textcolor{blue}{90.54} \\ 
 USG-Net\cite{r27USGNet} & \textcolor{blue}{84.52} & 99.02 & \textcolor{green}{85.44} & \textcolor{green}{86.78} & \textcolor{green}{86.88} \\ 
 \Xhline{1pt}  
 S\textsuperscript{3}TU-Net & \textcolor{red}{89.04} & \textcolor{blue}{99.53} & \textcolor{red}{90.70} & \textcolor{red}{90.73} & \textcolor{red}{93.70} \\ 
 \Xhline{1.1pt}  
\end{tabular}
\vspace{3mm}
\captionsetup{justification=centering}
\caption*{*Note: The red, blue and green rows represent the \textcolor{red}{1st}, \textcolor{blue}{2nd}, \textcolor{green}{3rd} places, respectively.}
\end{center}
\vspace{-0.2cm}
\end{table*}

\subsection{Comparative Experiments} 
To evaluate the segmentation performance of S\textsuperscript{3}TU-Net, this study compares it against several advanced and commonly used open-source methods for lung nodule segmentation, including YNet \cite{r22YNet}, UNet++ \cite{r23UNet++}, R2U-Net \cite{r26R2UNet}, Attention-UNet \cite{r24AttentionNet}, UNet3++ \cite{r28UNet3++}, USG-Net \cite{r27USGNet}, and SA-UNet \cite{r25SAUNet}. Table. \ref{table:performance_comparison_methods} presents the results on the LIDC-IDRI dataset. The top-performing model, USG-Net, achieved a DSC of 84.52\%, followed by SA-UNet with a DSC of 84.35\%. In contrast, the proposed S\textsuperscript{3}TU-Net model excels across all metrics, with a maximum DSC of 88.87\%, MIoU of 91.14\%, sensitivity of 93.48\%, and precision of 91.97\%, while its accuracy is comparable to that of other methods. Compared to the top-ranked models, USG-Net and SA-UNet, our model outperforms them in DSC by 4.52\% and 4.69\%, in mIoU by 5.26\% and 2.0\%, and in sensitivity by 6.82\% and 3.16\%, respectively.

To facilitate intuitive visualization, eight nodule images were randomly selected from the LIDC-IDRI dataset. Fig. \ref{fig:datu} presents the segmentation results of the two models with the highest DSC and the S\textsuperscript{3}TU-Net model. Compared with the classical methods and the more advanced models, our S\textsuperscript{3}TU-Net model, which combines SViT with spatial interaction mechanisms, consistently achieves smoother and more accurate segmentation results.

\section{Conclusion}\label{sec:con} 
Lung adenocarcinoma nodules in CT images often exhibit irregular and complex characteristics, posing significant challenges for accurate staging. Precise segmentation is essential for clinicians to focus on critical regions of interest and derive reliable diagnostic insights. While U-Net and its variants have demonstrated strong performance on conventional CT images, their generalization capability diminishes when handling complex adenocarcinoma nodules. To address this, we propose S\textsuperscript{3}TU-Net, a CNN-Transformer hybrid model that integrates structured convolution blocks, residual-based superpixel visual transformers (RM-SViT), and multi-dimensional spatial connectors (S\textsuperscript{2}-MLP Link) to enhance feature extraction, feature fusion, and global context understanding.
Key components such as DWF-Conv and D\textsuperscript{2}BR-Conv blocks improve global information representation and mitigate overfitting, while RM-SViT captures long-range dependencies efficiently through sparse correlation and multi-branch attention. Furthermore, the S\textsuperscript{2}-MLP Link module facilitates multi-scale feature transmission and minimizes information loss. On the LIDC-IDRI dataset, S\textsuperscript{3}TU-Net achieved a DSC of 89.04\%, precision of 90.73\%, IoU of 90.70\%, and sensitivity of 93.70\%. On the EPDB dataset, it achieved a DSC of 86.40\% and an accuracy of 98.53\%, demonstrating robust segmentation performance and strong generalization ability.

\section{Future Work}
Future research will focus on developing lightweight hybrid architectures to reduce computational overhead and improve inference speed, enabling real-time processing on resource-constrained devices, such as portable medical equipment. Additionally, we are extending the applicability of S\textsuperscript{3}TU-Net to other medical imaging modalities (e.g., ultrasound, MRI) and diverse disease segmentation tasks (e.g., dermatological conditions, prostate cancer). By leveraging its ability to capture both fine-grained local details and comprehensive global semantics, we aim to enhance segmentation accuracy and broaden its utility across medical domains.
Moreover, during our experiments, we identified the persistent challenge of medical image annotation. To address this, we plan to investigate self-supervised and semi-supervised learning approaches to improve generalization by utilizing minimally or unannotated data, reducing dependency on large-scale annotated datasets. These future directions aim to further enhance S\textsuperscript{3}TU-Net’s segmentation performance and versatility, paving the way for its seamless deployment and practical use in real-world clinical environments.

\bibliographystyle{IEEEtran}
\bibliography{ref} 

\begin{thebibliography}{10}
\providecommand{\url}[1]{#1}
\csname url@samestyle\endcsname
\providecommand{\newblock}{\relax}
\providecommand{\bibinfo}[2]{#2}
\providecommand{\BIBentrySTDinterwordspacing}{\spaceskip=0pt\relax}
\providecommand{\BIBentryALTinterwordstretchfactor}{4}
\providecommand{\BIBentryALTinterwordspacing}{\spaceskip=\fontdimen2\font plus
\BIBentryALTinterwordstretchfactor\fontdimen3\font minus \fontdimen4\font\relax}
\providecommand{\BIBforeignlanguage}[2]{{%
\expandafter\ifx\csname l@#1\endcsname\relax
\typeout{** WARNING: IEEEtran.bst: No hyphenation pattern has been}%
\typeout{** loaded for the language `#1'. Using the pattern for}%
\typeout{** the default language instead.}%
\else
\language=\csname l@#1\endcsname
\fi
#2}}
\providecommand{\BIBdecl}{\relax}
\BIBdecl

\bibitem{xin1}
H.-Y. Chen, H.-M. Wang, C.-H. Lin, R.~Yang, and C.-C. Lee, ``Lung cancer prediction using electronic claims records: A transformer-based approach,'' \emph{IEEE Journal of Biomedical and Health Informatics}, 2023.

\bibitem{xin2}
D.~Xiang, B.~Zhang, Y.~Lu, and S.~Deng, ``Modality-specific segmentation network for lung tumor segmentation in pet-ct images,'' \emph{IEEE Journal of Biomedical and Health Informatics}, vol.~27, no.~3, pp. 1237--1248, 2022.

\bibitem{xin3}
Z.~Li, J.~Zhang, T.~Tan, X.~Teng, X.~Sun, H.~Zhao, L.~Liu, Y.~Xiao, B.~Lee, Y.~Li \emph{et~al.}, ``Deep learning methods for lung cancer segmentation in whole-slide histopathology images—the acdc@ lunghp challenge 2019,'' \emph{IEEE Journal of Biomedical and Health Informatics}, vol.~25, no.~2, pp. 429--440, 2020.

\bibitem{lung2}
B.~D. Hutchinson, G.~S. Shroff, M.~T. Truong, and J.~P. Ko, ``Spectrum of lung adenocarcinoma,'' in \emph{Seminars in Ultrasound, CT and MRI}, vol.~40, no.~3.\hskip 1em plus 0.5em minus 0.4em\relax Elsevier, 2019, pp. 255--264.

\bibitem{lung3}
D.~J. Myers and J.~M. Wallen, ``Lung adenocarcinoma,'' in \emph{StatPearls [Internet]}.\hskip 1em plus 0.5em minus 0.4em\relax StatPearls Publishing, 2023.

\bibitem{r00}
A.~C. Borczuk, ``Prognostic considerations of the new world health organization classification of lung adenocarcinoma,'' \emph{European Respiratory Review}, vol.~25, no. 142, pp. 364--371, 2016.

\bibitem{r0}
W.~H. Organization \emph{et~al.}, ``Who classification of tumours of the lung, pleura, thymus and heart,'' \emph{WHO/IARC Classification of Tumours,}, vol.~7, 2015.

\bibitem{lung4}
X.~Shao, R.~Niu, Z.~Jiang, X.~Shao, and Y.~Wang, ``Role of pet/ct in management of early lung adenocarcinoma,'' \emph{American Journal of Roentgenology}, vol. 214, no.~2, pp. 437--445, 2020.

\bibitem{lung5}
J.~Cohen, E.~Reymond, A.~Jankowski, E.~Brambilla, F.~Arbib, S.~Lantuejoul, and G.~Ferretti, ``Lung adenocarcinomas: correlation of computed tomography and pathology findings,'' \emph{Diagnostic and interventional imaging}, vol.~97, no.~10, pp. 955--963, 2016.

\bibitem{xin4}
W.~Liu, X.~Liu, H.~Li, M.~Li, X.~Zhao, and Z.~Zhu, ``Integrating lung parenchyma segmentation and nodule detection with deep multi-task learning,'' \emph{IEEE Journal of Biomedical and Health Informatics}, vol.~25, no.~8, pp. 3073--3081, 2021.

\bibitem{shen2023advancing}
F.~Shen, H.~Ye, J.~Zhang, C.~Wang, X.~Han, and Y.~Wei, ``Advancing pose-guided image synthesis with progressive conditional diffusion models,'' in \emph{The Twelfth International Conference on Learning Representations}, 2023.

\bibitem{shen2024boosting}
F.~Shen, H.~Ye, S.~Liu, J.~Zhang, C.~Wang, X.~Han, and W.~Yang, ``Boosting consistency in story visualization with rich-contextual conditional diffusion models,'' \emph{arXiv preprint arXiv:2407.02482}, 2024.

\bibitem{lung7}
J.~Zhang, Y.~Xia, H.~Cui, and Y.~Zhang, ``Pulmonary nodule detection in medical images: a survey,'' \emph{Biomedical Signal Processing and Control}, vol.~43, pp. 138--147, 2018.

\bibitem{lung6}
R.~Paul, S.~H. Hawkins, Y.~Balagurunathan, M.~B. Schabath, R.~J. Gillies, L.~O. Hall, and D.~B. Goldgof, ``Deep feature transfer learning in combination with traditional features predicts survival among patients with lung adenocarcinoma,'' \emph{Tomography}, vol.~2, no.~4, p. 388, 2016.

\bibitem{hu2023bag}
J.~Hu, Z.~Huang, F.~Shen, D.~He, and Q.~Xian, ``A bag of tricks for fine-grained roof extraction,'' in \emph{IGARSS 2023-2023 IEEE International Geoscience and Remote Sensing Symposium}.\hskip 1em plus 0.5em minus 0.4em\relax IEEE, 2023.

\bibitem{old2}
H.~Liu, F.~Geng, Q.~Guo, C.~Zhang, and C.~Zhang, ``A fast weak-supervised pulmonary nodule segmentation method based on modified self-adaptive fcm algorithm,'' \emph{Soft Computing}, vol.~22, pp. 3983--3995, 2018.

\bibitem{old6}
H.~Xie, D.~Yang, N.~Sun, Z.~Chen, and Y.~Zhang, ``Automated pulmonary nodule detection in ct images using deep convolutional neural networks,'' \emph{Pattern recognition}, vol.~85, pp. 109--119, 2019.

\bibitem{old7}
S.~Jain, P.~Choudhari, and M.~Gour, ``Pulmonary lung nodule detection from computed tomography images using two-stage convolutional neural network,'' \emph{The Computer Journal}, vol.~66, no.~4, pp. 785--795, 2023.

\bibitem{hu2023robust}
J.~Hu, Z.~Huang, F.~Shen, D.~He, and Q.~Xian, ``A rubust method for roof extraction and height estimation,'' in \emph{IGARSS 2023-2023 IEEE International Geoscience and Remote Sensing Symposium}.\hskip 1em plus 0.5em minus 0.4em\relax IEEE, 2023.

\bibitem{old3}
M.~B. Tavakoli, M.~Orooji, M.~Teimouri, and R.~Shahabifar, ``Segmentation of the pulmonary nodule and the attached vessels in the ct scan of the chest using morphological features and topological skeleton of the nodule,'' \emph{IET Image Processing}, vol.~14, no.~8, pp. 1520--1528, 2020.

\bibitem{old4}
Y.~Zhang, F.-l. Chung, and S.~Wang, ``Clustering by transmission learning from data density to label manifold with statistical diffusion,'' \emph{Knowledge-Based Systems}, vol. 193, p. 105330, 2020.

\bibitem{r1}
O.~Ronneberger, P.~Fischer, and T.~Brox, ``U-net: Convolutional networks for biomedical image segmentation,'' in \emph{Medical image computing and computer-assisted intervention--MICCAI 2015: 18th international conference, Munich, Germany, October 5-9, 2015, proceedings, part III 18}.\hskip 1em plus 0.5em minus 0.4em\relax Springer, 2015, pp. 234--241.

\bibitem{unet2}
M.~Krithika Alias~AnbuDevi and K.~Suganthi, ``Review of semantic segmentation of medical images using modified architectures of unet,'' \emph{Diagnostics}, vol.~12, no.~12, p. 3064, 2022.

\bibitem{r3}
H.~Liu, H.~Cao, E.~Song, G.~Ma, X.~Xu, R.~Jin, Y.~Jin, and C.-C. Hung, ``A cascaded dual-pathway residual network for lung nodule segmentation in ct images,'' \emph{Physica Medica}, vol.~63, pp. 112--121, 2019.

\bibitem{r4}
D.~Lu, J.~Chu, R.~Zhao, Y.~Zhang, and G.~Tian, ``A novel deep learning network and its application for pulmonary nodule segmentation,'' \emph{Computational Intelligence and Neuroscience}, vol. 2022, no.~1, p. 7124902, 2022.

\bibitem{r5}
T.~Hou, J.~Zhao, Y.~Qiang, S.~Wang, and P.~Wang, ``Pulmonary nodules segmentation based on crf 3d-unet structure,'' \emph{Comput Eng Des}, vol.~41, no.~6, pp. 1663--1669, 2020.

\bibitem{old8}
X.~Arregui~Garc{\'\i}a, ``Vits vs. cnns for 3d medical image segmentation: Are transformers all you need?'' Master's thesis, 2023.

\bibitem{r6}
A.~Vaswani, ``Attention is all you need,'' \emph{Advances in Neural Information Processing Systems}, 2017.

\bibitem{yan2024enhancing}
K.~Yan, F.~Shen, and Z.~Li, ``Enhancing landslide segmentation with guide attention mechanism and fast fourier transformer,'' in \emph{International Conference on Intelligent Computing}.\hskip 1em plus 0.5em minus 0.4em\relax Springer, 2024, pp. 296--307.

\bibitem{r7}
A.~Dosovitskiy, ``An image is worth 16x16 words: Transformers for image recognition at scale,'' \emph{arXiv preprint arXiv:2010.11929}, 2020.

\bibitem{r8}
Z.~Liu, Y.~Lin, Y.~Cao, H.~Hu, Y.~Wei, Z.~Zhang, S.~Lin, and B.~Guo, ``Swin transformer: Hierarchical vision transformer using shifted windows,'' in \emph{Proceedings of the IEEE/CVF international conference on computer vision}, 2021, pp. 10\,012--10\,022.

\bibitem{r9dropout}
N.~Srivastava, G.~Hinton, A.~Krizhevsky, I.~Sutskever, and R.~Salakhutdinov, ``Dropout: a simple way to prevent neural networks from overfitting,'' \emph{The journal of machine learning research}, vol.~15, no.~1, pp. 1929--1958, 2014.

\bibitem{r10dropBlock}
G.~Ghiasi, T.-Y. Lin, and Q.~V. Le, ``Dropblock: A regularization method for convolutional networks,'' \emph{Advances in neural information processing systems}, vol.~31, 2018.

\bibitem{r11LKA}
M.-H. Guo, C.-Z. Lu, Z.-N. Liu, M.-M. Cheng, and S.-M. Hu, ``Visual attention network,'' \emph{Computational Visual Media}, vol.~9, no.~4, pp. 733--752, 2023.

\bibitem{r12FreeU}
C.~Si, Z.~Huang, Y.~Jiang, and Z.~Liu, ``Freeu: Free lunch in diffusion u-net,'' in \emph{Proceedings of the IEEE/CVF Conference on Computer Vision and Pattern Recognition}, 2024, pp. 4733--4743.

\bibitem{r13SENet}
J.~Hu, L.~Shen, and G.~Sun, ``Squeeze-and-excitation networks,'' in \emph{Proceedings of the IEEE conference on computer vision and pattern recognition}, 2018, pp. 7132--7141.

\bibitem{r15Ruan}
J.~Ruan, S.~Xiang, M.~Xie, T.~Liu, and Y.~Fu, ``Malunet: A multi-attention and light-weight unet for skin lesion segmentation,'' in \emph{2022 IEEE International Conference on Bioinformatics and Biomedicine (BIBM)}.\hskip 1em plus 0.5em minus 0.4em\relax IEEE, 2022, pp. 1150--1156.

\bibitem{r16Yu}
T.~Yu, X.~Li, Y.~Cai, M.~Sun, and P.~Li, ``S2-mlp: Spatial-shift mlp architecture for vision,'' in \emph{Proceedings of the IEEE/CVF winter conference on applications of computer vision}, 2022, pp. 297--306.

\bibitem{r17Tols}
I.~O. Tolstikhin, N.~Houlsby, A.~Kolesnikov, L.~Beyer, X.~Zhai, T.~Unterthiner, J.~Yung, A.~Steiner, D.~Keysers, J.~Uszkoreit \emph{et~al.}, ``Mlp-mixer: An all-mlp architecture for vision,'' \emph{Advances in neural information processing systems}, vol.~34, pp. 24\,261--24\,272, 2021.

\bibitem{r18MLP}
H.~Taud and J.-F. Mas, ``Multilayer perceptron (mlp),'' \emph{Geomatic approaches for modeling land change scenarios}, pp. 451--455, 2018.

\bibitem{r20LIDC}
H.~Zhang, C.~Wu, Z.~Zhang, Y.~Zhu, H.~Lin, Z.~Zhang, Y.~Sun, T.~He, J.~Mueller, R.~Manmatha \emph{et~al.}, ``Resnest: Split-attention networks,'' in \emph{Proceedings of the IEEE/CVF conference on computer vision and pattern recognition}, 2022, pp. 2736--2746.

\bibitem{r22YNet}
S.~Mehta, E.~Mercan, J.~Bartlett, D.~Weaver, J.~G. Elmore, and L.~Shapiro, ``Y-net: joint segmentation and classification for diagnosis of breast biopsy images,'' in \emph{Medical Image Computing and Computer Assisted Intervention--MICCAI 2018: 21st International Conference, Granada, Spain, September 16-20, 2018, Proceedings, Part II 11}.\hskip 1em plus 0.5em minus 0.4em\relax Springer, 2018, pp. 893--901.

\bibitem{r23UNet++}
Z.~Zhou, M.~M. Rahman~Siddiquee, N.~Tajbakhsh, and J.~Liang, ``Unet++: A nested u-net architecture for medical image segmentation,'' in \emph{Deep Learning in Medical Image Analysis and Multimodal Learning for Clinical Decision Support: 4th International Workshop, DLMIA 2018, and 8th International Workshop, ML-CDS 2018, Held in Conjunction with MICCAI 2018, Granada, Spain, September 20, 2018, Proceedings 4}.\hskip 1em plus 0.5em minus 0.4em\relax Springer, 2018, pp. 3--11.

\bibitem{r26R2UNet}
M.~Z. Alom, M.~Hasan, C.~Yakopcic, T.~M. Taha, and V.~K. Asari, ``Recurrent residual convolutional neural network based on u-net (r2u-net) for medical image segmentation,'' \emph{arXiv preprint arXiv:1802.06955}, 2018.

\bibitem{r24AttentionNet}
J.~Hu, Z.~Huang, F.~Shen, D.~He, and Q.~Xian, ``A bag of tricks for fine-grained roof extraction,'' 2023.

\bibitem{r28UNet3++}
H.~Huang, L.~Lin, R.~Tong, H.~Hu, Q.~Zhang, Y.~Iwamoto, X.~Han, Y.-W. Chen, and J.~Wu, ``Unet 3+: A full-scale connected unet for medical image segmentation,'' in \emph{ICASSP 2020-2020 IEEE international conference on acoustics, speech and signal processing (ICASSP)}.\hskip 1em plus 0.5em minus 0.4em\relax IEEE, 2020, pp. 1055--1059.

\bibitem{r25SAUNet}
C.~Guo, M.~Szemenyei, Y.~Yi, W.~Wang, B.~Chen, and C.~Fan, ``Sa-unet: Spatial attention u-net for retinal vessel segmentation,'' in \emph{2020 25th international conference on pattern recognition (ICPR)}.\hskip 1em plus 0.5em minus 0.4em\relax IEEE, 2021, pp. 1236--1242.

\bibitem{r27USGNet}
H.~Yang, L.~Shen, M.~Zhang, and Q.~Wang, ``Uncertainty-guided lung nodule segmentation with feature-aware attention,'' in \emph{International Conference on Medical Image Computing and Computer-Assisted Intervention}.\hskip 1em plus 0.5em minus 0.4em\relax Springer, 2022, pp. 44--54.

\end{thebibliography}
\end{document}